\documentclass[aps,
               prb,
               twocolumn,
               10pt,
               longbibliography,
               floatfix,
               nofootinbib,
               superscriptaddress]{revtex4-2}

\usepackage{cancel}
\usepackage{physics}
\usepackage[english]{babel}
\usepackage[utf8]{inputenc}
\usepackage{amssymb}
\usepackage{amsmath}
\usepackage{multirow}
\usepackage{float}
\usepackage[pdftex]{graphicx}
\usepackage{xspace}
\usepackage{dsfont}
\usepackage{soul}
\usepackage{bm}
\usepackage[normalem]{ulem}
\usepackage{hyperref}
\hypersetup{
    colorlinks,%
    citecolor=blue,%
    filecolor=blue,%
    linkcolor=blue,%
    urlcolor=blue
}
\usepackage[usenames, dvipsnames]{color}

\newcommand{\FBA}[0]{1BA\xspace}

\graphicspath{{.}{Figuras/}} 

\begin{document}

\title{Phase transitions and scale invariance in topological Anderson insulators}
\author{Bryan D. Assun\c{c}\~ao}
\affiliation{Instituto de F\'{i}sica, Universidade Federal de Uberl\^{a}ndia, Uberl\^{a}ndia, Minas Gerais 38400-902, Brazil}

\author{Gerson J. Ferreira}
\affiliation{Instituto de F\'{i}sica, Universidade Federal de Uberl\^{a}ndia, Uberl\^{a}ndia, Minas Gerais 38400-902, Brazil}

\author{Caio H. Lewenkopf}
\affiliation{Instituto de Física, Universidade Federal Fluminense, 24210-346 Niterói, Rio de Janeiro, Brazil}

\date{\today}
    
\begin{abstract}
We investigate disorder-driven transitions between trivial and topological insulator (TI) phases in two-dimensional (2D) systems. 
Our study primarily focuses on the Bernevig-Hughes-Zhang (BHZ) model with Anderson disorder, while other standard 2DTI models exhibit equivalent features.
The analysis is based on the local Chern marker (LCM), a local quantity that allows for the characterization of topological transitions in finite and disordered systems. 
Our simulations indicate that disorder-driven trivial to topological insulator transitions are nicely characterized by $\mathcal{C}_0$, the \textit{disorder-averaged} LCM near the central cell of the system. 
We show that $\mathcal{C}_0$ is characterized by a single-parameter scaling, namely, $\mathcal{C}_0(M, W, L) \equiv \mathcal{C}_0(z)$, with $z = [W^\mu-W_c^\mu(M)]L$, where $M$ is the Dirac mass, $W$ is the disorder strength, and $L$ is the system size, while 
$W_c(M) \propto \sqrt{M}$ 
and $\mu \approx 2$ stand for the critical disorder strength and the critical exponent, respectively. 
Our numerical results are in agreement with a theoretical prediction based on a first-order Born approximation analysis. 
These observations lead us to speculate that the universal scaling function we have found is rather general for amorphous and disorder-driven topological phase transitions.
\end{abstract}

\maketitle

%%%%%%%%%%%%%%%%%%%%%%%%%%%%%%%%%%%%%%%%%%%%%%%%%%
{\it Introduction.--} 
%%%%%%%%%%%%%%%%%%%%%%%%%%%%%%%%%%%%%%%%%%%%%%%%%%
Topological insulators are a fascinating new class of materials characterized by unique electronic properties
\cite{Hasan2010,Qi2011,Culcer2020}.
Notable among their attributes is the appearance of symmetry-protected spin-polarized edge (or surface) states that contrast with the insulating bulk region \cite{Kane2005, Bernevig2006}. 
These materials find diverse applications in spintronics, encompassing spin-to-charge conversion devices \cite{Sahu2023}, memory read-out, and field-effect transistors in quantum computing \cite{Nikoofard2022}, and they demonstrate promising potential in preparing electrodes for photodetectors \cite{Wang2020} and other applications \cite{He2019, Mazumder2021, Breunig2021}.

Since the unique features of such materials manifest only in their topological phase, there is great interest in a deeper understanding of their robustness. 
The theory predicts that smooth deformations of the system's band structure do not change its topological properties, provided the topological gap does not close \cite{Hasan2010, Qi2011}. 
In pristine systems, topological invariants \cite{Kane2005, Kaufmann2016} have been used to characterize the topological phase and to study the trivial to topological transitions caused by magnetic fields \cite{Tom2021}, electric fields \cite{You2021}, mechanical strain \cite{Zheng2021}, temperature variations \cite{Antonius2016}, doping, and interactions \cite{Melo2023SSH}.
The understanding of disordered systems is less clear since the standard topological invariants rely on translation symmetry.

A systematic study of topological phase transitions \cite{Chen2016} has put forward a scaling procedure for inversion-symmetric topological insulators, successfully identifying critical and fixed points under scaling of the curvature function. Similar results have been obtained for other topological systems, such as in static and periodically driven Kitaev chains \cite{Molignini2018} and in higher-order band crossings \cite{Chen2019}.
Topological phase transitions have also been studied using the local Chern marker (LCM) as a function of a mass parameter in the Haldane model \cite{Caio2019}. 

\begin{figure}[H]
    \centering
    \includegraphics[width=\columnwidth]{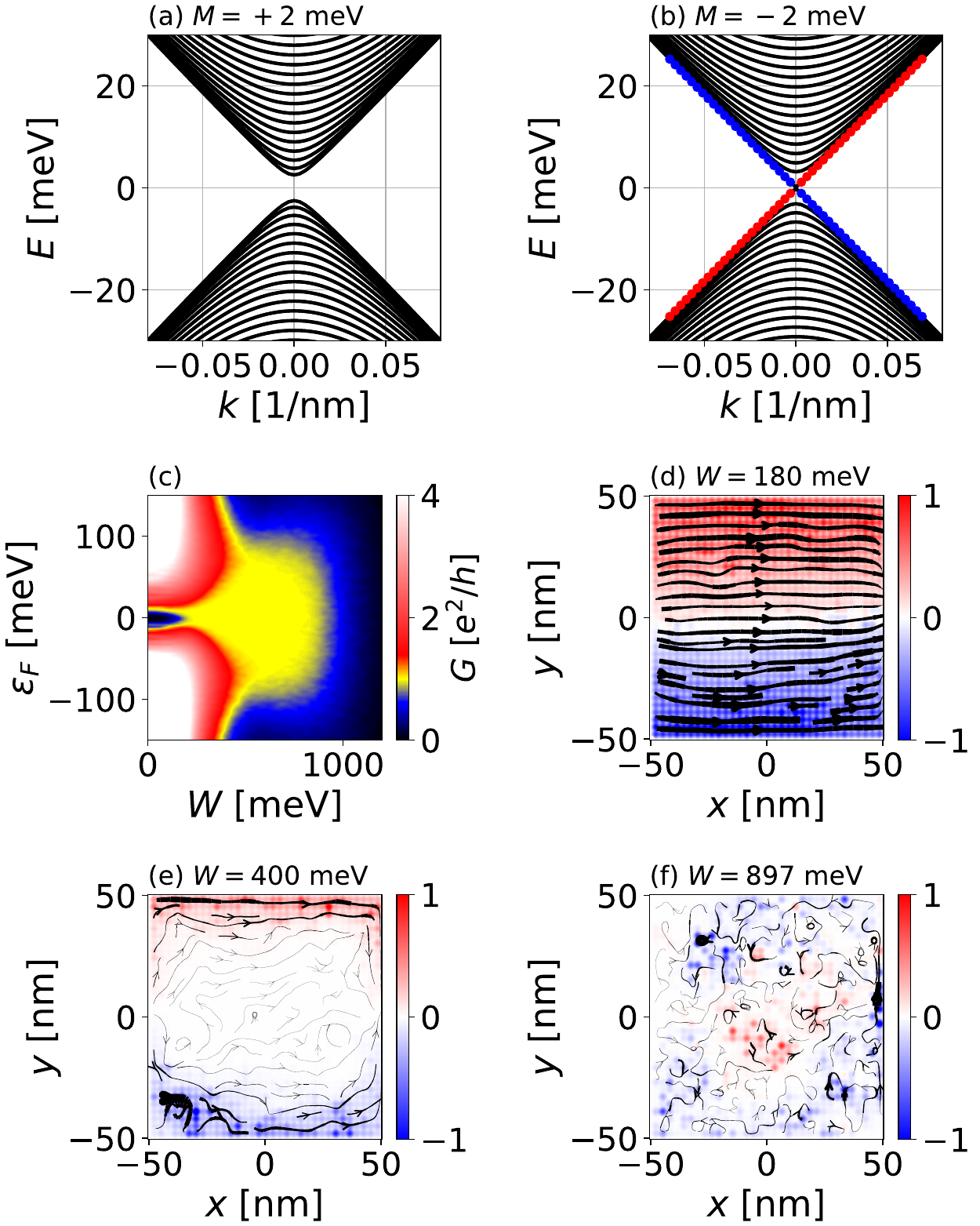}
    \caption{
    (a,b) 
    Band structure of the BHZ Hamiltonian for a pristine ribbon of width $L_y = 600$~nm with
    (a) $M = 2$~meV, and (b) $M=-2$~meV.
    (c) Conductance $G$ averaged over $N = 100$ disorder realizations, as a function of the Fermi energy $\varepsilon_F$ and the disorder strength $W$.
    The yellow region highlights the $G = G_0$ quantized plateau corresponding to the TAI phase.
    (d)-(f) Current density (black lines and arrows) and spin polarization density (color coded with spin up in red, and spin down in blue) for a single-disorder realization with $L_x = L_y = 100$~nm and (d) $W = 180$~meV, (e) $W = 400$~meV, and (f) $W = 897$~meV.}
    \label{fig:BandsG}
\end{figure}

Of particular interest are disorder-driven, such as the topological Anderson insulators (TAI) \cite{Li2009, Groth2009}, and amorphous \cite{Agarwala2017, Costa2019, Corbae2023b, Regis2023} topological systems, where one has to rely on local topological invariants, such as the LCM \cite{Resta2011, Bianco2011, dOrnellas2022}. 
Recent studies have extended the two-dimensional LCM put forward by Bianco-Resta \cite{Bianco2011} to odd dimensions  \cite{Sykes2021OddDim}, defined a local Chern marker for periodic systems \cite{bau2023local, *StraWBerryPy} and finite temperatures \cite{Molignini2023LCMOpacity}, and established conditions for their robustness against disorder \cite{oliveira2024robustness}, opening new paths for future investigations.
Various studies have identified TAI systems, including disordered HgTe quantum wells \cite{Jiang2009, Prodan2011, Song2012, Xu2012}, Haldane models and Kane-Mele systems \cite{Zhang2019, Orth2016}, bilayer crystals \cite{Liu2023}, three-dimensional systems \cite{Shindou2009}, four-dimensional artificial lattices \cite{RuiChen2023TAI4D}, 
higher-order TAI in Sierpi\'nski lattices \cite{HuanChen2023Sierpinski}, half-filled Haldane models with extended Hubbard interactions \cite{gonçalves2018haldane}, the Haldane model with binary disorder \cite{PhysRevB.109.125145}, and others \cite{Guo2010, Titum2015}, where nontrivial phases arise in areas of the phase diagram where the clean limit is topologically trivial.
The disordered topological phase can also be classified by investigating how the conductance approaches the quantization plateau \cite{Girschik2013, Vu2022}. 
Experimental realizations include heterostructures \cite{Liu2020} and disordered atomic wires \cite{Meier2018}, among others \cite{Sttzer2018, ZangenehNejad2020}. 
However, a study of scaling and critical exponents in disorder-driven topological phase transitions is still lacking.

In this Letter, we introduce the {\it disorder-averaged} local Chern marker at the center of the system $\mathcal{C}_0$ to characterize trivial to topological phase transitions driven by the disorder. 
For finite-size systems, $\mathcal{C}_0$ varies smoothly as a function of the disorder strength $W$, as defined by the Anderson disorder model, allowing for a scale invariance investigation with the system size $L$. 
Our findings are based on the study of three seminal models for two-dimensional (2D) topological insulators (TIs), namely, the Bernevig-Hughes-Zhang (BHZ) \cite{BHZ2006}, Haldane \cite{Haldane1988}, and Kane-Mele \cite{KaneMele2005} models. Here we focus on the BHZ model, while in order to avoid unnecessary repetitions, equivalent results for the Haldane and Kane-Mele models are shown in the Supplemental Materials \cite{supmat}.
In all cases, our simulations indicate that $\mathcal{C}_0$ scales as $\mathcal{C}_0(W, L) \equiv \mathcal{C}_0(z)$, with $z = (W^\mu-W_c^\mu)L$, where $W_c$ is the critical disorder strength and $\mu\approx 2$ is the critical exponent in the trivial to TAI phase transition.
These results are further supported by a theory based on the scaling analysis of first-order Born approximation (1BA) results. 
These observations lead us to speculate that our findings are universal for disorder-driven topological phase transitions.

%%%%%%%%%%%%%%%%%%%%%%%%%%%%%%%%%%%%
{\it Model and methods.--} 
%%%%%%%%%%%%%%%%%%%%%%%%%%%%%%%%%%%%
%
The BHZ Hamiltonian \cite{BHZ2006} is one of the standard models to describe 2D TIs in the AII symmetry class of the Altland-Zirnbauer tenfold way classification \cite{Altland1997, Chiu2016}. 
Considering only the majority spin-up block, the BHZ Hamiltonian reads as
\begin{align}
    H = C - Dk^2 + A \bm{k}\cdot\bm{\sigma} + (M-Bk^2)\sigma_z,
    \label{eq:BHZ}
\end{align}
where $\bm{k} = (k_x, k_y)$ is the quasimomentum, and $\bm{\sigma} = (\sigma_x, \sigma_y, \sigma_z)$ are the Pauli matrices acting on the $E_1/H_1$ space of subbands of HgTe quantum wells \cite{BHZ2006}.
As standard, we use $A = 365$~meV nm, $B=686$~meV nm$^2$, and the mass term $M = 2$~meV, unless otherwise specified. 
For simplicity, we choose to set $C=D=0$ to keep the energy spectrum particle-hole symmetric, and to allow for a direct comparison with the Haldane \cite{Haldane1988} and Kane-Mele \cite{KaneMele2005} models (see Supplemental Material \cite{supmat} for details).
For the numerical calculations, $\bm{k} = -i\bm{\nabla}$ is discretized into the square lattice with lattice parameter $a = 3$~nm, defining finite systems of variable sizes $L_x \times L_y$.
The Anderson disorder is introduced as uncorrelated random onsite energies uniformly distributed in the range $[-W/2, W/2]$.
We compute the Landauer conductance $G$ and the current densities by standard methods \cite{Datta1995, kwant}, considering this region coupled to semi-infinite pristine leads of width $L_y$. 

Figures \ref{fig:BandsG}(a) and \ref{fig:BandsG}(b) show the band structure of the BHZ Hamiltonian in a ribbon geometry of width $L_y = 600$~nm and periodic along $x$ for $M=+2$~meV and $M=-2$~meV, respectively. 
In both cases, we consider pristine systems ($W = 0$) to emphasize that, in the trivial insulator case, there is a band gap of $\approx 2M = 4$~meV, while in the nontrivial regime ($M<0$) the gap is filled with topologically protected helical edge states, which are responsible for a quantized conductance of $G_0 = e^2/h$. 
As shown in Ref.~\cite{Groth2009}, and revised in the Supplemental Materials \cite{supmat}, within the \FBA the Anderson disorder ($W\neq 0$) renormalizes the mass term as
\begin{align}
    M \rightarrow \widetilde{M} \approx M - \alpha(M) W^2.
    \label{eq:Meff}
\end{align}
This important result reveals that disorder can drive the system from a trivial regime ($M=2$~meV, $W=0$) into a TAI regime with an effective mass of $\widetilde{M} < 0$. 
Figure~\ref{fig:BandsG}(c) shows the conductance $G$ for the disordered BHZ model as a function of the Fermi energy $\varepsilon_F$ and the disorder intensity $W$. 
The large parameter space region of the quantized conductance $G = G_0$ (in yellow) characterizes a TAI phase.
Figures \ref{fig:BandsG}(d) and \ref{fig:BandsG}(e) illustrate the current and spin densities for a smaller system ($L=100$~nm) for different disorder intensities and $\varepsilon_F = 0$. 
For small $W$, as shown in Fig.~\ref{fig:BandsG}(d), the system is already near a TAI phase and the current and spin densities show a tendency to form helical edge states. 
Figure \ref{fig:BandsG}(e) shows that as the disorder strength $W$ is increased, the system reaches the TAI phase and the helical edge states become well-defined.
For even larger $W$ values, comparable with the bandwidth, the topological protection is destroyed and the states become localized, as shown by Fig.~\ref{fig:BandsG}(f), indicating the onset of Anderson localization.

%%%%%%%%%%%%%%%%%%%%%%%%%%%%%%%%%%
{\it Local Chern marker.--} 
%%%%%%%%%%%%%%%%%%%%%%%%%%%%%%%%%%
% 
As discussed in Ref.~\cite{Li2009}, in the trivial regime the conductance is dominated by bulk transport, and the scaled conductance $\sigma = G L_x/L_y$ decays with $W$ as a universal function, independent of the ribbon width $L_y$. 
In contrast, at the TAI phase, $G = G_0$ forms a quantized plateau [see Fig.~\ref{fig:BandsG}(c)]. 
Therefore, instead of using $G$, we find it convenient to use local markers to investigate the normal-TAI transition, its scale invariance properties, and critical exponents.
As shown in Ref.~\cite{Bianco2011}, the local Chern marker can be written as \cite{LCMderivation}
\begin{align}
    \mathcal{C}(\bm{R}) &= \dfrac{4\pi}{A_c} \Im 
    \sum_\beta 
    \int_{\bm{R}}d^2r
    \bra{\bm{r},\beta} 
    \hat{U} \hat{P} \hat{x} \hat{P} \hat{y} \hat{P} 
    \ket{\bm{r},\beta},
    \label{eq:LCM}
\end{align}
where $\hat{U}$ is defined below, the integral is taken over a unit cell of area $A_c$ centered at $\bm{R}$, the sum runs over the states $\beta$ that define the basis for the model Hamiltonian (e.g., $E_1/H_1$ orbitals for the BHZ model), $\hat{x}$ and $\hat{y}$ are the position operators, and $\hat{P}$ is the projector over occupied states, defined as
\begin{align}
    \hat{P} = \sum_\ell^{\epsilon_\ell < \varepsilon_F} \ketbra{\psi_\ell}.
    \label{eq:ProjP}
\end{align}
Here $\epsilon_\ell$ and  $\ket{\psi_\ell}$ are the eigenenergies and eigenstates of the Hamiltonian $H$ for a finite, nonperiodic system, and the Fermi energy is set to $\varepsilon_F = 0$ hereafter.
It has been shown that 
$\mathcal{C}(\bm{R})$ recovers the quantized bulk Chern number $C$ as one restores the lattice periodicity for clean systems \cite{Sykes2021OddDim},
while for finite systems $\mathcal{C}(\bm{R}) \approx C$ for $\bm{R} \approx 0$, near the center of the sample \cite{Bianco2011, Caio2019}. 
For the BHZ model, defined by Eq.~\eqref{eq:BHZ}, and for the Haldane model \cite{supmat}, we consider $\hat{U} = 1$. 
On the other hand, for the spinful Kane-Mele model \cite{supmat}, we set $\hat{U} = \sigma_z$ (Pauli matrix in spin space), such that $\mathcal{C}(\bm{R})$ represents a local spin-Chern number \cite{SpinChernNumber}. In this case, for a finite Rashba coupling, discussed in Ref.~\cite{supmat}, the spin-Chern number is not quantized, but it remains a reliable witness for nontrivial topology, as shown in Refs.~\cite{Ezawa2013, Ezawa2014, Zhu2019}.

\begin{figure}[t]
    \centering
    \includegraphics[width=\columnwidth]{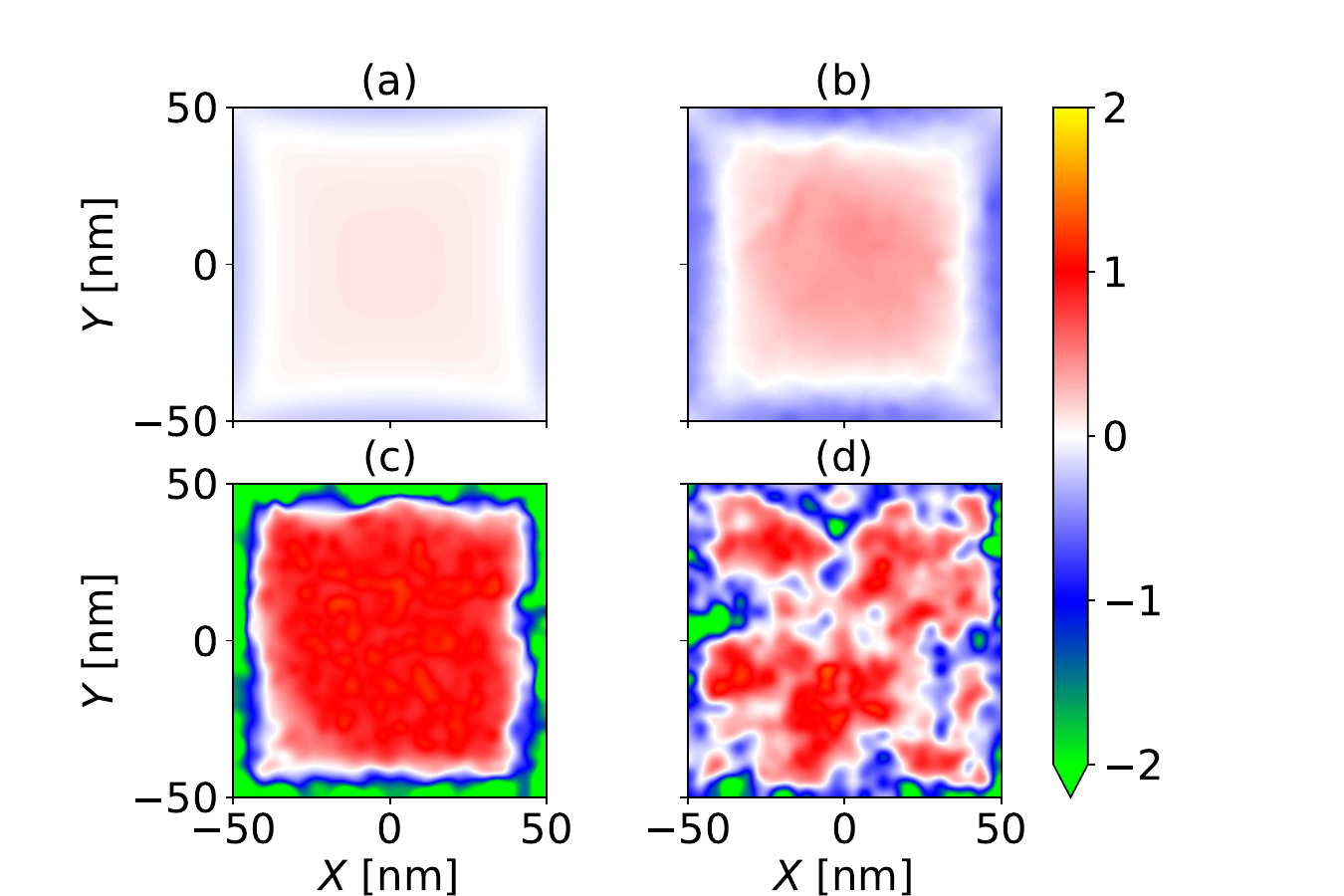}
    \caption{Local Chern marker $\mathcal{C}(\bm{R})$ for a representative single-disorder realization as a function of $\bm{R} = (X, Y)$ for a system of size $L_{x} = L_{y} = 100$~nm, $M = +2$~meV, and disorder strengths
    (a) $W = 25$~meV, (b) $W = 180$~meV, (c) $W = 400$~meV, and (d) $W = 897$~meV.
    Panels (b) to (d) correspond to the current and spin densities shown in Figs.~\ref{fig:BandsG}(e) and \ref{fig:BandsG}(f).}
    \label{fig:mapC}
\end{figure}

In the presence of Anderson disorder, the local marker $\mathcal{C}(\bm{R})$ acquires fluctuations. 
Consequently, the analysis of the LCM requires a statistical approach. 
Here, we focus on the average $\mathcal{C}(\bm{R})$, which can be taken over disorder realizations and/or spatial averages. 
For large systems, this can be an intensive numerical task. Therefore, we have implemented two approaches. 
First, to calculate full $\mathcal{C}(\bm{R})$ maps as a function of $\bm{R}$ for a single-disorder sample (e.g., in Fig.~\ref{fig:mapC}), we use a \textit{brute force} approach, where the finite-size system is fully diagonalized, $H\ket{\psi_\ell} = \epsilon_\ell \ket{\psi_\ell}$, such that the operators $\hat{P}$, $\hat{x}$, and $\hat{y}$ can be explicitly calculated to yield $\mathcal{C}(\bm{R})$ from Eq.~\eqref{eq:LCM}.
In the second approach, we define the disorder-averaged local marker $\mathcal{C}_0$ as the average of $\mathcal{C}(\bm{R}=0)$ 
over many disorder realizations, namely, $\mathcal{C}_0 \equiv \langle \mathcal{C}(\bm{R}=0)\rangle$.
We compute $\mathcal{C}_0$ employing the efficient implementation developed in Ref.~\cite{Varjas2020} based on the kernel polynomial method (KPM) \cite{Weisse2006}. 
The KPM allows us to estimate $\hat{P}\ket{\varphi}$ for an arbitrary state $\ket{\varphi}$ in terms of an efficient expansion of Chebyshev polynomials 
(details about the KPM implementation are presented in Ref.~\cite{supmat}).

%%%%%%%%%%%%%%%%%%%%%%%%%%%%%%%%%%%%
%{\it Results.--}
%%%%%%%%%%%%%%%%%%%%%%%%%%%%%%%%%%%%
%
Figure~\ref{fig:mapC} shows $\mathcal{C}(\bm{R})$ as a function of $\bm{R} = (X, Y)$ for a single-disorder realization for increasing disorder strengths $W$. 
Figure~\ref{fig:mapC}(a) corresponds to a small $W$ for which the system is in the trivial regime, which is characterized by a small LCM near the sample center, that is,  $\mathcal{C}({\bm R}\approx 0) \approx 0$.
Figure~\ref{fig:mapC}(b) considers a moderate disorder intensity $W$, yielding $\mathcal{C}({\bm R}\approx 0) \approx 0.7$ and thus approaching a TAI regime, which is fully achieved for larger $W$, as shown in Fig.~\ref{fig:mapC}(c), where $\mathcal{C}({\bm R}\approx 0) \approx 1.0$. 
Notice that Figs.~\ref{fig:mapC}(b) and \ref{fig:mapC}(c) correspond to the same values of $W$ of Figs.~\ref{fig:BandsG}(d) and \ref{fig:BandsG}(e).
Interestingly, up to Fig.~\ref{fig:mapC}(c) or Fig.~\ref{fig:BandsG}(e), we find that the Anderson disorder has driven the system into a TAI phase, but there are no clear signs of Anderson localization, since the maps are overall smooth, apart from small fluctuations that cause some blurriness. 
However, for even larger $W$ [see Fig.~\ref{fig:mapC}(d) or Fig.~\ref{fig:BandsG}(f)], Anderson localization emerges as strong fluctuations over the full map $\mathcal{C}(\bm{R})$.
In all cases, we see that $\mathcal{C}(\bm{R})$ takes negative values at the edges, which fulfills the constraint  $\sum_{\bm{R}}\mathcal{C}(\bm{R}) = 0$ \cite{Bianco2011}. Indeed, this is an undesirable consequence of the approximations that take place in the derivation of the local marker \cite{LCMderivation}, which, for now, we leave as an issue to be discussed in future works. 

\begin{figure}[t]
    \centering
    \includegraphics[width=\columnwidth]{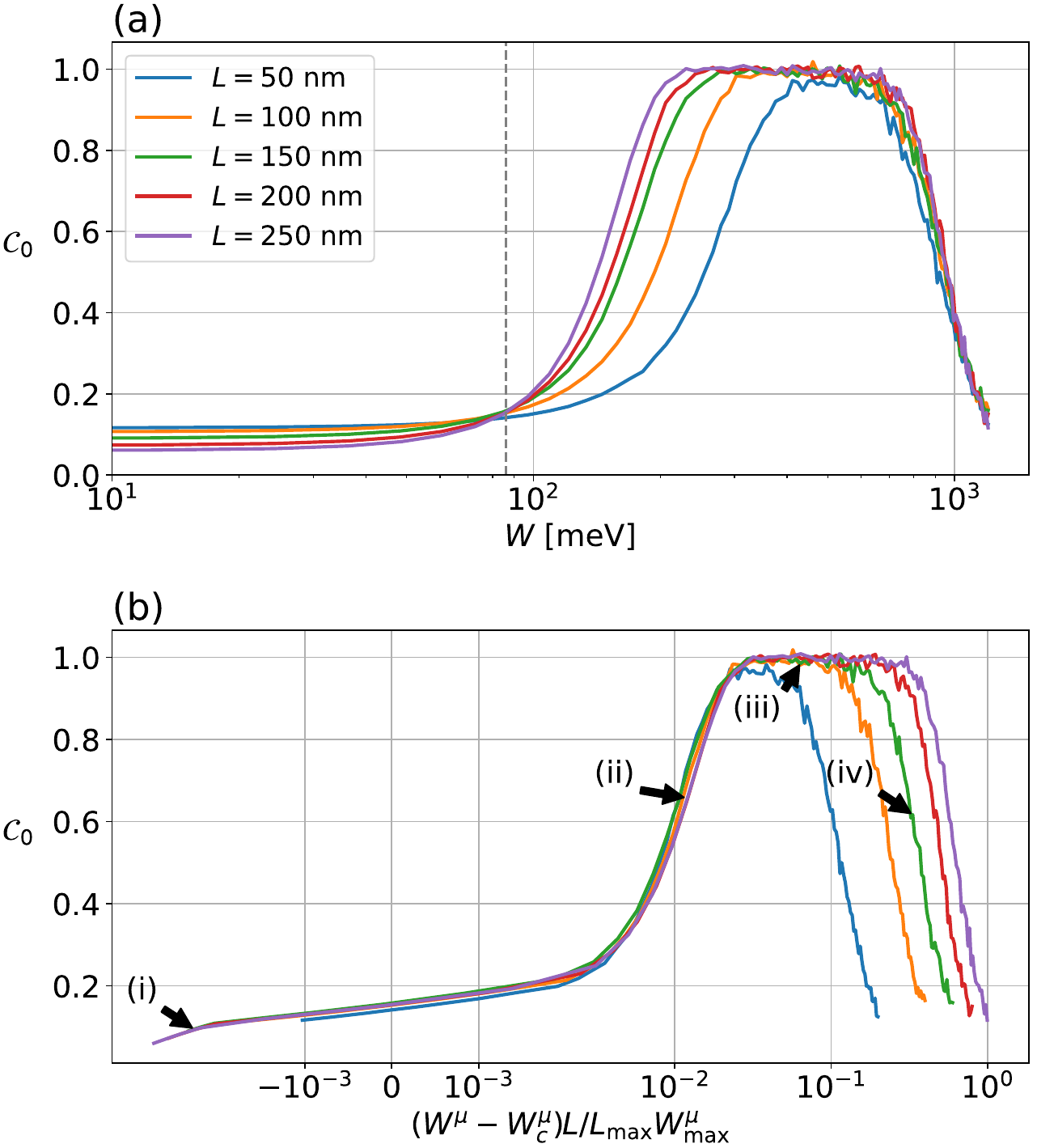}
    \caption{(a) Disorder-averaged local Chern marker (400 disorder realizations and 2000 Chebyshev moments in the KPM expansion) as a function of $W$ for different $L_x=L_y=L$ showing a trivial-TAI transition, $\mathcal{C}_0 \rightarrow 0$ to 1, followed by the trivial Anderson localization as $W$ becomes larger than the bandwidth. (b) Optimal scaling of the marker showing that all curves in panel (a) fall into a universal trend for the trivial-TAI transition region upon the scaling $W \rightarrow z = (W^\mu-W_c^\mu) L/W_{\rm max}^\mu L_{\rm max}$, with $\mu \approx 2$ and $W_c \approx 86$~meV.
    The points indicated by the arrows (i) to (iv) correspond to the ${\cal C}({\bm R})$ color maps shown in Fig.~\ref{fig:mapC}(a) to \ref{fig:mapC}(d), respectively.
    }
    \label{fig:LCM0}
\end{figure}

%%%%%%%%%%%%%%%%%%%%%%%%%%%%%%%%%%%%
{\it Phase transition and scaling.--}
%%%%%%%%%%%%%%%%%%%%%%%%%%%%%%%%%%%%
% 
Figure~\ref{fig:LCM0} shows the disorder-averaged local Chern marker $\mathcal{C}_0$ 
as a function of $W$
for different system sizes in a square geometry, $L_x = L_y \equiv L$, and a {\it fixed} $M = +2$~meV.
As $W$ increases, all cases first display a transition from a trivial regime with $\mathcal{C}_0 \approx 0.1$ to a nontrivial regime characterized by $\mathcal{C}_0 \approx 1$.
For very large values of $W$, exceeding the bandwidth ($E_{\rm max} \sim |B|k_{\rm max}^2 = 8|B|/a^2 \approx 610$~meV for $a=3$~nm) of the pristine BHZ Hamiltonian, $\mathcal{C}_0$ collapses back to $\mathcal{C}_0 \rightarrow 0$. 
This is the transition seen from Fig.~\ref{fig:mapC}(c) to Fig.~\ref{fig:mapC}(d), which is due to the emergence of Anderson localization. 
Such behavior is not surprising, since in such a strong-disorder regime, it hardly makes sense to consider the system as topological.
Figure \ref{fig:LCM0}(a) clearly shows that
the second transition, $\mathcal{C}_0 = 1 \rightarrow 0$, does not depend on $L$.
In fact, the transition to a trivial Anderson insulator 
is governed by an energy scale defined by the bandwidth \cite{Anderson1958}, which is only influenced by $L$ for very small samples.
Thus, for moderate $L$, both the bandwidth and this transition are expected to be independent of $L$.
In contrast, the first transition, $\mathcal{C}_0 \approx 0.1 \rightarrow 1$, is $L$ dependent. This dependence is what we analyze next.

We obtain a single-parameter universal scaling function for the trivial-TAI phase transition by expressing ${\cal C}_0$ in terms of the parameter $z$, which combines the disorder strength $W$ and the system size $L$ according to
\begin{align}
    \mathcal{C}_0(W, L) \rightarrow \mathcal{C}_0(z),
    \label{eq:CofZ} 
\end{align}
with
\begin{align}
    z = \dfrac{W^\mu-W_c^\mu}{W_{\rm max}^\mu} \cdot \dfrac{L}{L_{\rm max}},
    \label{eq:WtoZ}
\end{align}
as shown in Fig.~\ref{fig:LCM0}(b). 
Here, $W_c$ is the critical point where the curves corresponding to systems with different sizes $L$ cross, and $\mu$ is the critical exponent. 
To make $z$ dimensionless, we introduce $L_{\rm max} = 250$~nm, the largest simulated system size, and $W_{\rm max} = 1200$~meV, the largest considered disorder strength, in the denominators of $z$.
We find numerically that the optimal scaling in Fig.~\ref{fig:LCM0}(b) occurs for $\mu \approx 2$ and $W_c \approx 86$~meV [dashed line in Fig.~\ref{fig:LCM0}(a)]. 
Next, to understand these values for $\mu$ and $W_c$, we analyze the phase diagram and scaling properties of the $\mathcal{C}_0$ as a function of both the mass $M$ and the disorder intensity $W$.

\begin{figure}[t!]
    \centering
    \includegraphics[width=\columnwidth]{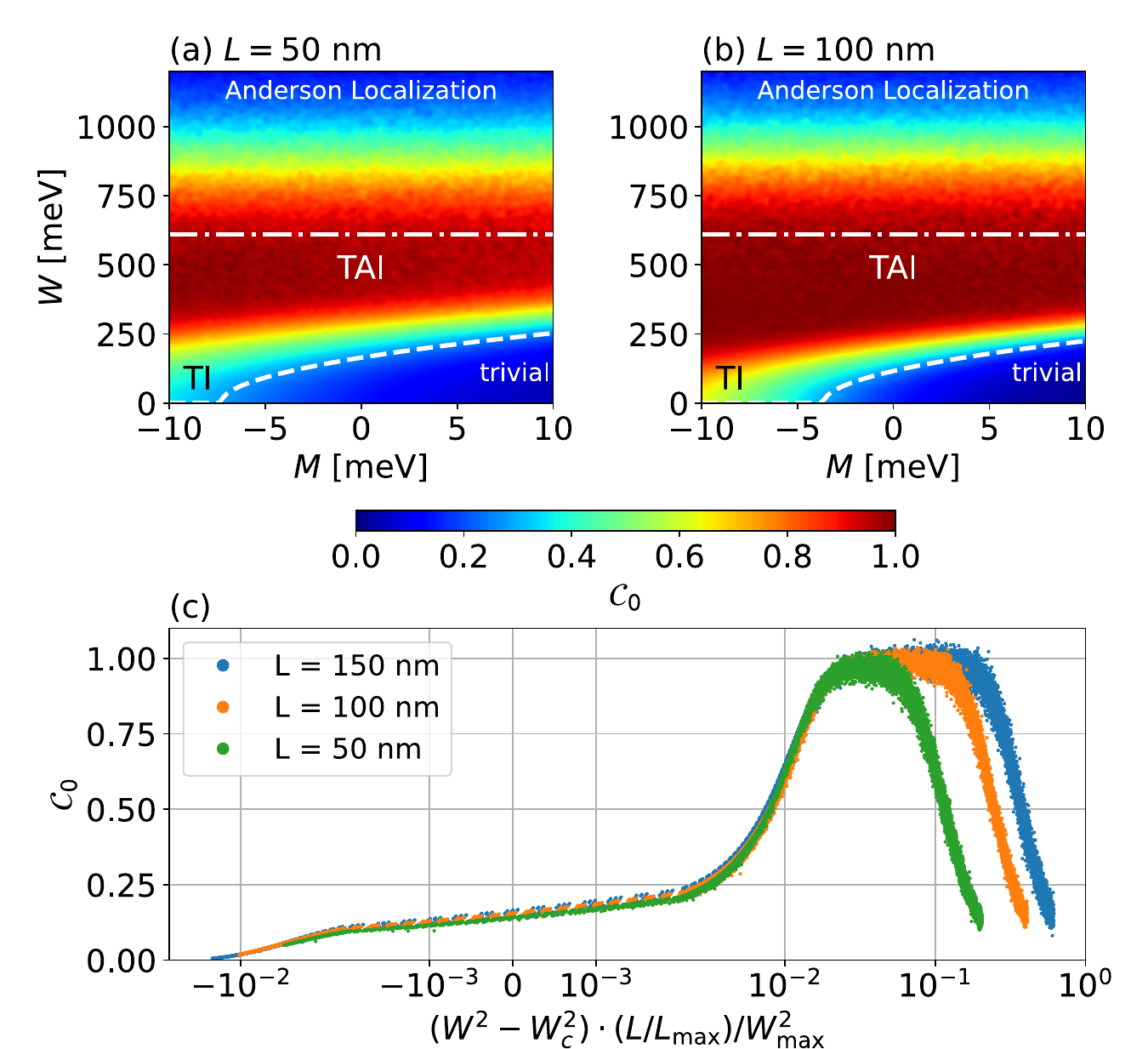}
    \caption{(a, b) Phase diagrams calculated using the local Chern marker $\mathcal{C}_0$ as a function of the BHZ mass $M$ and the disorder strength $W$ for (a) $L = 50$~nm and (b) 100~nm.
    In all cases, we consider 200 disorder realizations and 1000 Chebyshev moments on the KPM expansion.
    The dashed white line marks the phase boundary between the trivial and topological phases (TI and TAI) obtained from the gapless condition $A/|\widetilde{M}| \ll L$.
    The dot-dashed line marks the boundary where $W$ matches the bandwidth of the clean system, above which the Anderson localization phase takes place.
    (c) All data from panels (a) and (b), and additional data for a system of $L=150$~nm, fall into a universal curve for the trivial to TAI phase transition upon the scaling $(M, W, L)\rightarrow z = [W^\mu - W_c^\mu(M)] L / W^\mu_{\rm max} L_{\rm max}$.}
    \label{fig:PhaseDiag}
\end{figure}

%%%%%%%%%%%%%%%%%%%%%%%%%%%%%%%%%%%%
{\it Phase diagram.--}
%%%%%%%%%%%%%%%%%%%%%%%%%%%%%%%%%%%%
% 
The local Chern marker can be used to define a topological phase diagram. For the BHZ model, we now consider $\mathcal{C}_0$ in terms of the mass $M$ and the disorder strength $W$, as shown in Fig.~\ref{fig:PhaseDiag}, for two system sizes. 
In both cases the color code labels the value of the local marker $\mathcal{C}_0$ 
calculated for 400 disorder realizations. 
The phase diagrams in Figs.~\ref{fig:PhaseDiag}(a) and \ref{fig:PhaseDiag}(b) show the following phases: trivial, TI, TAI, and Anderson localization (AL). 
As discussed above, the AL phase occurs when the disorder strength exceeds the bandwidth $W > E_{\rm max} \approx 610$~meV. 

We recall that, for $W=0$, the bulk Chern number indicates that a TI phase occurs for $M < 0$, whereas Figs.~\ref{fig:PhaseDiag}(a) and \ref{fig:PhaseDiag}(b) show that the trivial to TI transition, characterized by the local marker $\mathcal{C}_0$, depends upon the finite system size $L$. 
In a pristine 2D TI, the topological edge states' are gapless and do not hybridize, provided the system size $L$ is much larger than the edge states penetration depth $\xi = A/|M|$, i.e., for $A/|M| \ll L$ (with $M < 0$ and $W=0$).
% 
% |Mt| >> A/L
% |M-a.W²| >> A/L
% |M-a.W²| = A/L
% M-a.W² = -A/L
% W² = (M+A/L)/a
% 
We extend this gapless condition to the disordered case by making $M \rightarrow \widetilde{M}$ to obtain an expression for the phase boundary using $\widetilde{M}$ from \FBA in Eq.~\eqref{eq:Meff}. 
Hence, from the gapless condition $A/|\widetilde{M}| \ll L$ and assuming $\widetilde{M} \leq 0$, we obtain
\begin{align}
    W_{\rm gapless}(M) = 
    \sqrt{\dfrac{M+A/L}{\alpha(M)}}
    \approx\sqrt{\dfrac{M+A/L}{\alpha_0}}.
\end{align}
Here and in what follows we use that the first Born approximation gives $\alpha(M) \approx \alpha_0$ \cite{supmat}. 
In Figs.~\ref{fig:PhaseDiag}(a) and \ref{fig:PhaseDiag}(b) the dashed lines mark the phase boundary given by $W_{\rm gapless}(M)$ for an optimal $\alpha_0 \approx 1/3700$~meV$^{-1}$, which matches remarkably well the phase boundary between the trivial and topological phases (TI and TAI). 

The data shown in Figs.~\ref{fig:PhaseDiag}(a) and \ref{fig:PhaseDiag}(b) for all $M$ and $W$, and additional data for $L=150$~nm (not shown), can be combined into a single universal function that characterizes the trivial to TAI phase transition shown in Fig.~\ref{fig:PhaseDiag}(c). 
The scaling used in this case is a direct extension of the one presented in Eqs.~\eqref{eq:CofZ} and \eqref{eq:WtoZ}, where now $W_c \rightarrow W_c(M)$ depends upon $M$. 
To obtain an expression for $W_c(M)$ and to understand the value $\mu \approx 2$ of the critical parameters, let us first consider the clean regime ($W = 0$), to find the scaling law for the trivial to TI phase transition as a function of $M$. 
Here, we consider an argument similar to the one used in Ref.~\cite{Caio2019} for the Haldane model, as follows. 
In the topological regime, the only relevant length scales are $\xi = A/|M|$ and $L$, as introduced above. 
Therefore, one can expect that the topological phase transition scaling with the system size $L$ should be a universal function of $L/\xi \propto ML$ (assuming $A$ is constant). Thus, we justify that
\begin{align}
    \mathcal{C}_0(M, L) \equiv \mathcal{C}_0(ML).
    \label{eq:CML}
\end{align}
Next, to extend this scaling to finite $W$, we replace $M\rightarrow\widetilde{M}$ using the \FBA expression from Eq.~\eqref{eq:Meff} with $\alpha(M) \approx \alpha_0$, such that the universal function argument now reads
\begin{align}
    \label{eq:ML}
    \widetilde{M}L = (M -  \alpha_0 W^2)L
    = -\alpha_0(W^2 - W_c^2)L,
\end{align}
where $W_c^2 = M/\alpha_0$ is the critical point and marks the phase boundary in the thermodynamic limit ($L/\xi\rightarrow \infty$), where $\widetilde{M} = 0$ and $W_{\rm gapless} = W_c$. Indeed, for $M = 2$~meV, this expression yields $W_c \approx 86$~meV, thus matching the value found in Fig.~\ref{fig:LCM0}. Moreover, apart from constant factors, the expression for $\widetilde{M}L$ in Eq.~\eqref{eq:ML} matches $z$ given by Eq.~\eqref{eq:WtoZ} with $\mu=2$, in agreement with the previously numerically obtained value. 

%%%%%%%%%%%%%%%%%%%%%%%%%%%%%%%%%%%%
{\it Conclusions.--}
%%%%%%%%%%%%%%%%%%%%%%%%%%%%%%%%%%%%
% 
We have investigated the scale invariance of the local Chern marker $\mathcal{C}_0$ applied to TAIs. The smooth profile of $\mathcal{C}_0$ throughout the phase transition allows us to characterize the scale invariance and obtain critical parameters, and universal scaling functions. 
Our analysis uses the disordered BHZ model as a canonical example of TI and TAI phases. 
The simplicity of the model allows for an analytical justification of the obtained numerical results. 
We like to stress that the disordered Haldane \cite{Haldane1988} and Kane-Mele \cite{KaneMele2005} models exhibit an identical scaling and universal behavior, as we have shown in the Supplemental Material \cite{supmat}.
This observation supports the statement that our findings are quite generic. 
Indeed, as long as the first-order Born approximation remains sufficiently accurate, we expect that the universal function will scale with $\sim W^2 L$, while deviations from this behavior should occur only beyond the validity range of the \FBA. 
We believe that our findings shed light on an approach to investigating scaling properties of topological phase transitions for disordered and amorphous systems. 

{\it Note added in proof.}
The independent study \cite{mildner2023topological} has considerable overlap with our work.
% Recently, Ref.~\cite{mildner2023topological} -- which has a large overlap with our study -- appeared as a preprint in the arXiv.

%%%%%%%%%%%%%%%%%%%%%%%%%%%%%%%%%%%%%
{\it Acknowledgments.--}
% \begin{acknowledgments}
    This work was supported by the Brazilian funding agencies CNPq, CAPES, FAPERJ, and FAPEMIG (Grant No. PPM-00798-18).
    The authors acknowledge useful conversations with C.~Beenakker, A.~Akhmerov, and C.~P.~Orth.
% \end{acknowledgments}

\bibliography{main}

\end{document}

\title{Supplemental Material to 
\texorpdfstring{\\}{}
``Phase transitions and scale invariance in Topological Anderson Insulators"}
\author{Bryan D. Assun\c{c}\~ao}
\affiliation{Instituto de F\'{i}sica, Universidade Federal de Uberl\^{a}ndia, Uberl\^{a}ndia, MG 38400-902, Brazil}

\author{Gerson J. Ferreira}
\affiliation{Instituto de F\'{i}sica, Universidade Federal de Uberl\^{a}ndia, Uberl\^{a}ndia, MG 38400-902, Brazil}

\author{Caio H. Lewenkopf}
\affiliation{Instituto de Física, Universidade Federal Fluminense, 24210-346 Niterói, RJ, Brazil}

\date{\today}
\maketitle

\tableofcontents

\section{Summary}

Here, we present the following additional numerical and analytical materials: 
(\ref{sec:selfenergy}) Details on the computation of the self-energy, within the first order Born approximation, due to random impurity scattering in the BHZ model. First, we consider an approximate analytical expression for the continuum model, and next, we present numerical results for the BHZ model regularized into a square lattice.
(\ref{sec:KPM}) We show how the kernel polynomial method (KPM) is used to efficiently calculate the local Chern marker, and we discuss the convergence of the KPM expansion.
(\ref{sec:othermodels}) We show results for the Haldane \cite{Haldane1988} and the Kane-Mele \cite{KaneMele2005} models, that complement the study of the BHZ model presented in the main text. As referred to in the main text, the analysis of the latter corroborates our findings. These results are presented as additional material to avoid unnecessary repetitions. 

%%%%%%%%%%%%%%%%%%%%%%%%%%%%%%%%%%%%%%
\section{Impurity scattering self-energy}
\label{sec:selfenergy}

The clean system is described by the BHZ Hamiltonian
\begin{align}
    H_0 &= A(k_x \sigma_x + k_y \sigma_y) + (M+Bk^2)\sigma_z,
\end{align}
where we have already set $C=D=0$ for simplicity. 
A random impurity perturbation potential $V$ is added to the full Hamiltonian $H = H_0 + V$, such that the Dyson equation reads $G(\bm{k}) = G_0(\bm{k}) + G_0(\bm{k}) \Sigma G(\bm{k})$. 
Here, $G_0(\bm{k}) = [\varepsilon_F - H_0 + i0^+]^{-1}$ is the bare Green's function, $\Sigma = \expval{V} + \expval{VG_0V} + \cdots$ is the self-energy due to the impurities, the $\expval{\cdot}$ refers to the impurity self-averaging, and $G(\bm{k})$ is the full Green's function. The latter is diagonal in $\bm{k}$ as the impurity self-averaging restores translation invariance \cite{BruusFlensberg}. Within the first-order Born approximation, the self-energy reads 
\begin{align}
    \Sigma_{\rm 1BA}
    &= \dfrac{W^2}{12} 
    \left(\dfrac{a}{2\pi}\right)^2
    \int_{\rm BZ} G_0(\bm{k}) d^2k,
\end{align}
which is independent of $\bm{k}$. The factor $(a/2\pi)^2$ is the inverse area of the Brillouin zone (BZ) related to the square lattice with lattice parameter $a$. 
The on-site random impurity potential $W_i$ is uniformly distributed within the range $[-W/2, W/2]$, such that $\expval{W_i} = 0$ and $\expval{W_i W_j} = \frac{1}{12}W^2 \delta_{i,j}$. 
Here, we are interested in the real (Hermitian) part of $\Sigma_{\rm 1BA}$, which leads to chemical potential and mass renormalizations. The chemical potential renormalization can be neglected, since it only shifts the energy reference. Hence, hereafter we consider only the mass renormalization, that is given by
\begin{equation}
    \label{eq:mass-renomalization-def}
    \widetilde{M} = M + \Re\Sigma_z = M + \alpha(M) W^2,
\end{equation}
with
\begin{equation}
    \label{eq:alpha-def}
    \alpha(M) = \dfrac{1}{12} 
    \left(\dfrac{a}{2\pi}\right)^2
    \int_{\rm BZ} \dfrac{1}{2} \Tr\left[\sigma_z G_0(\bm{k})\right] d^2k.
\end{equation}

For the sake of simplicity, we take the chemical potential $\varepsilon_F = 0$. 
Consequently, $\varepsilon_F$ lies within the gap between the BHZ bands and, hence, the self-energy should be identically real, since its imaginary part is proportional to the density of states at the chemical potential, \textit{i.e.} $\Im\Sigma_{\rm 1BA} \propto D(\varepsilon_F) = 0$. 

In Ref.~\cite{Groth2009} the authors arrive at an approximate analytical expression for the renormalization parameters by considering only the logarithmically divergent part of the $\Sigma_{\rm 1BA}$ integral.
Within this approximation, Eq.~\eqref{eq:alpha-def} leads to $\alpha(M) \approx (a^2/48\pi B)\ln(\pi^4 B^2/(M^2)a^4)$, for $C=D=0$.
However, this approximation yields unphysical divergencies for $|M| = \varepsilon_F = 0$. 
In what follows, we show that a careful analysis of $\alpha(M)$ gives a smooth and nearly constant function of $M$.

To obtain an analytical expression for the mass renormalization defined in Eqs.~\eqref{eq:mass-renomalization-def} and \eqref{eq:alpha-def}, and its correction beyond the logarithmically divergent part, we express $\bm{k} = (k\cos\theta, k\sin\theta)$ in polar coordinates, replace the integral over the BZ by a integral over a disk of radius $k_{\rm cut} = \pi/a$, assuming that the most relevant contributions arise from the region near $\bm{k} = 0$. 
Later we will contrast this approximation with a numerical integral over the full BZ. 
For now, we notice that our approximation significantly simplifies the problem, as it yields the $\theta$-integral as
\begin{align}
    \dfrac{1}{2}\Tr\left[\sigma_z \int_0^{2\pi} G_0(\bm{k}) d\theta\right] &= \dfrac{-2\pi(M+Bk^2)}{(Ak)^2 + (M+Bk^2)^2}.
\end{align}
In turn, for $k \le k_{\rm cut}$ the $k$-integral gives
\begin{widetext}
\begin{equation}
\label{eq:alpha}
    \alpha(M) \approx
    \dfrac{a^2}{96\pi B}\log
    \left[
        \dfrac{(Ak_{\rm cut})^2 + (M+Bk_{\rm cut}^2)^2}{M^2}
    \right]
    \\+
    \dfrac{a^2 A}{48 \pi B\sqrt{A^2+4BM}}
    \arctanh
    \left[
        \dfrac{A^2+2B(M+Bk^2)}{A\sqrt{A^2+4BM}}
    \right]\Bigg|_0^{k_{\rm cut}}.
\end{equation}
\end{widetext}
By considering only the logarithmically divergent part and leading order in $k_{\rm cut} \rightarrow \infty$, we obtain $\alpha(M) \approx (a^2/96\pi B)\ln(\pi^4 B^2/(M^2)a^4)$, which matches the expression for $\alpha(M)$ present in Ref.~\cite{Groth2009} (see above), apart from an overall factor 1/2. 
While this logarithmically divergent part diverges at $M=0$, the full expression given by Eq.~\eqref{eq:alpha} is well behaved at this limit and yields $\lim_{M\rightarrow 0} \alpha(M) = (a^2/48\pi B) \ln[1+(B k_{\rm cut} /A)^2]$, which shows that the $\arctanh$ contribution cannot be neglected for small $M$. 

To verify the precision of our approximate analytical expression for $\alpha(M)$, we numerically integrate Eq.~\eqref{eq:alpha-def}. 
For that purpose, we consider the regularization of the BHZ Hamiltonian in a square lattice of lattice parameter $a$, such that the $k$-powers in $H_0$ become $k_x \rightarrow \frac{1}{a}\sin(k_x a)$, $k_x^2 \rightarrow \frac{2}{a^2}[1-\cos(k_x a)]$, and equivalent expressions for $k_y$ and $k_y^2$. 
Hence, the BZ is defined by the square region set by $|k_x| \leq \pi/a$ and $|k_y| \leq \pi/a$.

\begin{figure}
    \centering
    \includegraphics[width=\columnwidth]{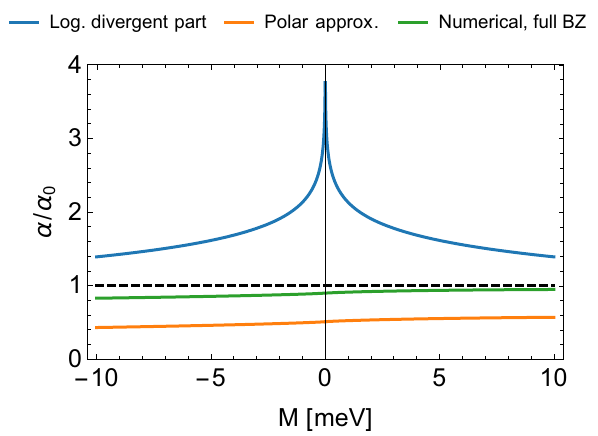}
    \caption{Comparison of $\alpha(M)/\alpha_0$ as a function of the bare mass $M$ between the analytical approximation of Eq.~\eqref{eq:alpha} (orange), its logarithmically divergent part (blue), and the numerical integration over a square lattice (green). Here $\alpha_0 = 1/3700$~meV$^{-1}$ is the reference value.}
    \label{fig:alpha}
\end{figure}

The analytical expression given by Eq.~\eqref{eq:alpha}, its logarithmically divergent part, and the numerical integration results are compared in Fig.~\ref{fig:alpha}, where we plot $\alpha/\alpha_0$ as a function of the bare mass $M$. 
Here $\alpha_0 = 1/3700$~meV$^{-1}$ is the optimal value used in the main text. 
The comparison shows that $\alpha(M)$ obtained using the polar cutoff approximation,  Eq.~\eqref{eq:alpha}, matches well the smooth behavior of the numerics except for a rigid shift that is likely due to the missing regions in polar integration of the BZ. 
Furthermore, the numerical results fall close to the constant value $\alpha_0 = 1/3700$~meV$^{-1}$, indicating that $\alpha$ depends very weekly on $M$.
In the main text, we use this value for the scaling analysis.

It is important to emphasize that all calculations shown here are done for bulk systems and within the first-order Born approximation. 
In contrast, the results presented in the paper consider finite systems and numerical averaging over many realizations of the random impurity ensemble. 
In other words, the assumption that $\alpha \approx \alpha_0$ (used in the main text), which is well established by the numerics (see Fig.~\ref{fig:alpha}), does not account for eventual finite size corrections.  

%%%%%%%%%%%%%%%%%%%%%%%%%%%%%%%%%%%%%%
\section{The kernel polynomial method}
\label{sec:KPM}

The KPM expansion \cite{Weisse2006} for the projector operator over occupied states is given by
\begin{align}
    P(\varepsilon, H) = \theta(\varepsilon - H)
    = \sum_{m=0}^{M} g_m \mu_m(\varepsilon) T_m (\widetilde{H}),
\end{align}
where $T_m(H)$ are the Chebyshev polynomials, $g_m$ are the Jackson kernel coefficients \cite{Weisse2006}, $\widetilde{H} = (H-b)/a$ is the Hamiltonian $H$ shifted by $b$ and renormalized by $a$ such that the energy spectrum lies in the range $[-1, +1]$ to match the Chebyshev polynomial domain, and $\mu_m(\varepsilon)$ are the moments of the expansion, which read as \cite{Varjas2020}
\begin{align}
    \mu_m(\varepsilon) 
    &= \frac{2}{1+ \delta_{m, 0}} \int_{-1}^{1} \frac{\theta(\varepsilon - E) T_{m}(E)}{\pi\sqrt{1 - E^{2}}} dE,
    \nonumber \\
    &= 
    \begin{cases}
        1 - \frac{\arccos(\varepsilon)}{\pi}, & \text{for } m=0,
        \\
        \dfrac{-2 \sin[m \arccos(\varepsilon)]}{m \pi}, & \text{for } m \neq 0.
    \end{cases}
\end{align}
In the definition of $\mu_m(\varepsilon)$, the denominator $\pi\sqrt{1-E^2}$ is the weight function for the inner product of Chebyshev polynomials of the first kind, and the prefactor before the integral appears due to the orthogonality $\braket{T_n}{T_m} = \frac{1}{2}(1+\delta_{m,0})\delta_{n,m}$.

In practice, rather than evaluating $\hat{P}$ itself, we need to calculate the action $\hat{P}\ket{r}$ on a generic state vector $\ket{v}$. Namely,
\begin{align}
    P(\varepsilon, H)\ket{v} = \sum_{m=0}^{M} g_m \mu_m(\varepsilon) \ket{v_m}.
\end{align}
The vectors $\ket{v_m} = T_m (\widetilde{H})\ket{v}$ can be efficiently obtained via the recursion relation of Chebyshev polynomials, yielding
\begin{align}
    \ket{v_0} &= \ket{v},
    \\
    \ket{v_1} &= \widetilde{H}\ket{v_0},
    \\
    \ket{v_m} &= 2\widetilde{H}\ket{v_{m-1}} - \ket{v_{m-2}}.
\end{align}

For the LCM, we need to compute matrix elements of the type $\bra{v} \hat{U} \hat{P} \hat{x} \hat{P} \hat{y} \hat{P} \ket{v}$, as shown in Eq.~3 of the main text. This can be acomplished using the KPM recipe above by parts as
\begin{align}
    \ket{\phi_1} &= \hat{P}\ket{v}, % P.r
    \\
    \ket{\phi_2} &= \hat{y}\ket{\phi_1} = \hat{y}\hat{P}\ket{v}, % yP.r
    \\
    \ket{\phi_3} &= \hat{x}\ket{\phi_1} = \hat{x}\hat{P}\ket{v}, % xP.r
    \\
    \ket{\phi_4} &= \hat{P}\ket{\phi_3} = \hat{P}\hat{x}\hat{P}\ket{v}, % PxP.r
\end{align}
where $\ket{\phi_1}$ and $\ket{\phi_4}$ are calculated via KPM and the others are straightforward. These can be combined to form $\bra{v} \hat{U} \hat{P} \hat{x} \hat{P} \hat{y} \hat{P} \ket{v} = \bra{\phi_4} \hat{U} \ket{\phi_2}$, which assumes that $\hat{U}$ commutes with $\hat{P}$, $\hat{x}$, and $\hat{y}$ \cite{Varjas2020}.

\subsection{Convergence analysis}

For all results presented in this Letter, we have rigorously tested the convergence of the LCM calculated via KPM, as explained in this section.

\begin{figure}[h]
    \centering
    \includegraphics[width=\columnwidth]{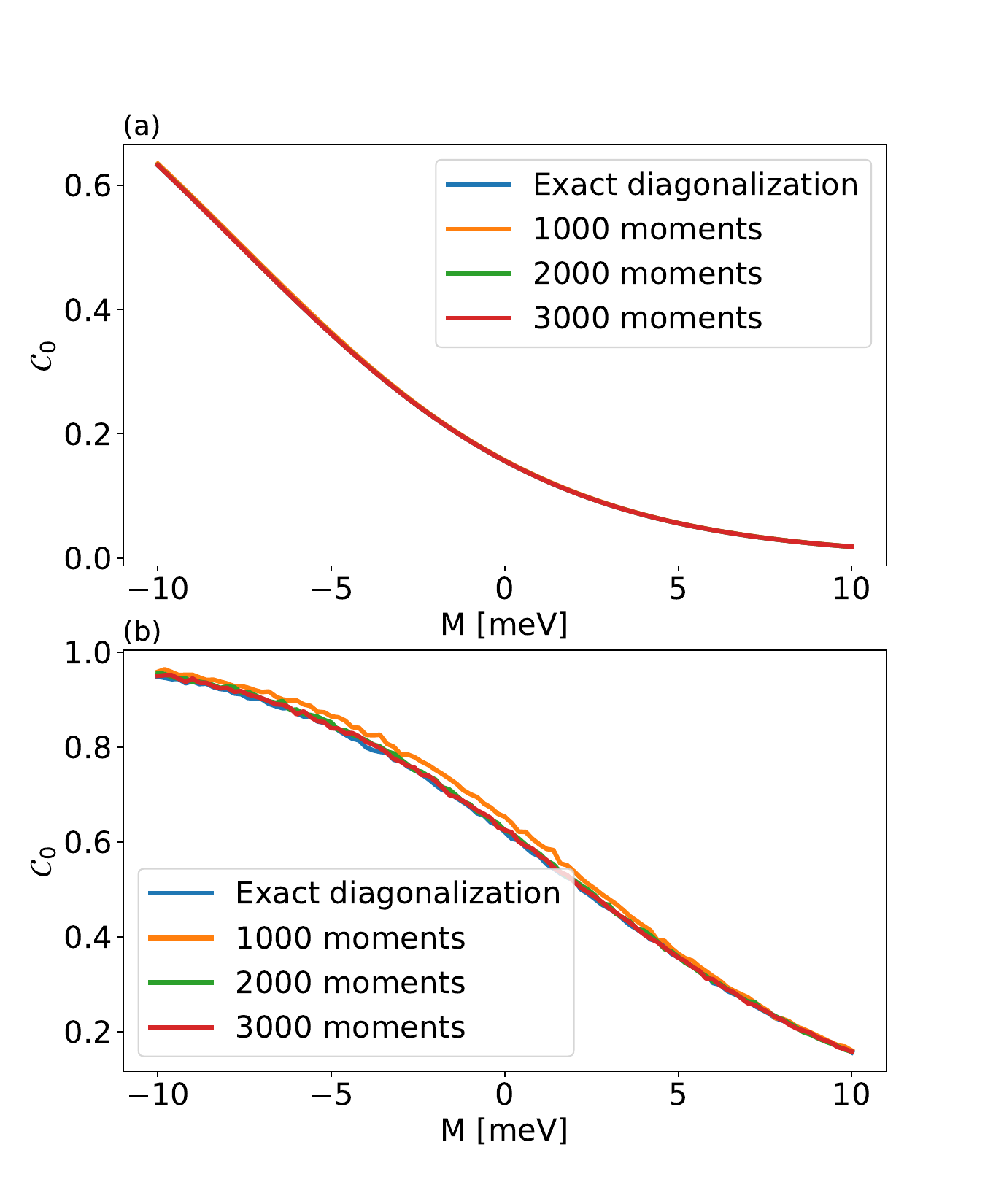}
    \caption{Comparison of the local Chern marker calculated via exact diagonalization (\textit{brute force}, blue line) and via the KPM method with different number of Chebyshev moments in the expansion. Here we consider the BHZ model with $L = 100$~nm, and vary the mass $M$ for fixed (a) $W=0$, and (b) $W = 200$~meV. 
    (a) 
    For $W=0$, all lines fall on top of each other, and the discrepancy between the KPM and exact diagonalization data is less than $1\%$ in all cases.
    (b) 
    For $W \neq 0$ the error is typically around $2$ to $3\%$ and at most about $5\%$ in the worst cases.}
    \label{fig:KPM}
\end{figure}

To illustrate the convergence analysis, we consider the BHZ model on a flake of size $L = 100$~nm. In Fig.~\ref{fig:KPM} we compare the LCM $\mathcal{C}_0$ calculated via exact diagonalization (\textit{brute force}) and via KPM with different number of Chebyshev moments. Fig.~\ref{fig:KPM}(a) show the results for the clean system with $W=0$, while Fig.~\ref{fig:KPM}(b) presents the data for $W = 200$~meV averaged over 400 disorder realizations. In both cases we vary the mass $M$ in the same range presented in Fig.~4 of the main text. 
In all cases, the data shows that the LCM for 1000 moments is close to convergence, and from 2000 to 3000 moments there no significant changes.

%%%%%%%%%%%%%%%%%%%%%%%%%%%%%%%%%%%%%%
\section{Results for other models}
\label{sec:othermodels}

The main text analyzes the topological phase transitions due to Anderson disorder in the BHZ model \cite{BHZ2006}. Here, we extend this analysis to other standard topological insulator models, namely, the Haldane \cite{Haldane1988} and the Kane-Mele \cite{Kane2005} models.
We find that these models show the same scaling behavior discussed in the main text.

%%%%%%%%%%%%%%%%%%%%%%%%%%%%%%%%%%%%%%
\subsection{Haldane model}
\label{sec:Haldane}

The Hamiltonian of the Haldane model \cite{Haldane1988} reads as 
\begin{multline}
    H_{\rm H} = -t_1 \sum_{\expval{ij}} (c_i^\dagger c_j + {\rm h.c.}) - t_2 \sum_{\expval{\expval{ij}}} (e^{i\phi_{ij}} c_i^\dagger c_j + {\rm H.c.}) 
    \\
    + M \sum_{i \in A} c_i^\dagger c_i^{} -M \sum_{i \in B} c_i^\dagger c_i^{},
\end{multline}
where $i$ and $j$ label the sites of a honeycomb lattice, a hexagonal lattice with a two-atom basis, labeled by $A$ and $B$. 
The first neighbor hopping term $t_1 \equiv t$ sets the energy units and we take $t_2=t/3$ for the second neighbor hopping. 
Here, we analyze the model dependence on the mass $M$ and the phase $\phi$ over the ranges $|M| \leq 2t$ and $0 \leq \phi \leq 2\pi$. 
Additionally, we consider Anderson's onsite disorder, such that the total Hamiltonian reads
\begin{equation}
    H = H_{\rm H} + \sum_i \varepsilon_i c_i^\dagger c_i^{},
\end{equation}
where the onsite energies $\{\varepsilon_i\}$ are randomly chosen from a uniform distribution within the interval $[-W/2, W/2]$.

\begin{figure*}
    \includegraphics[width=\textwidth]{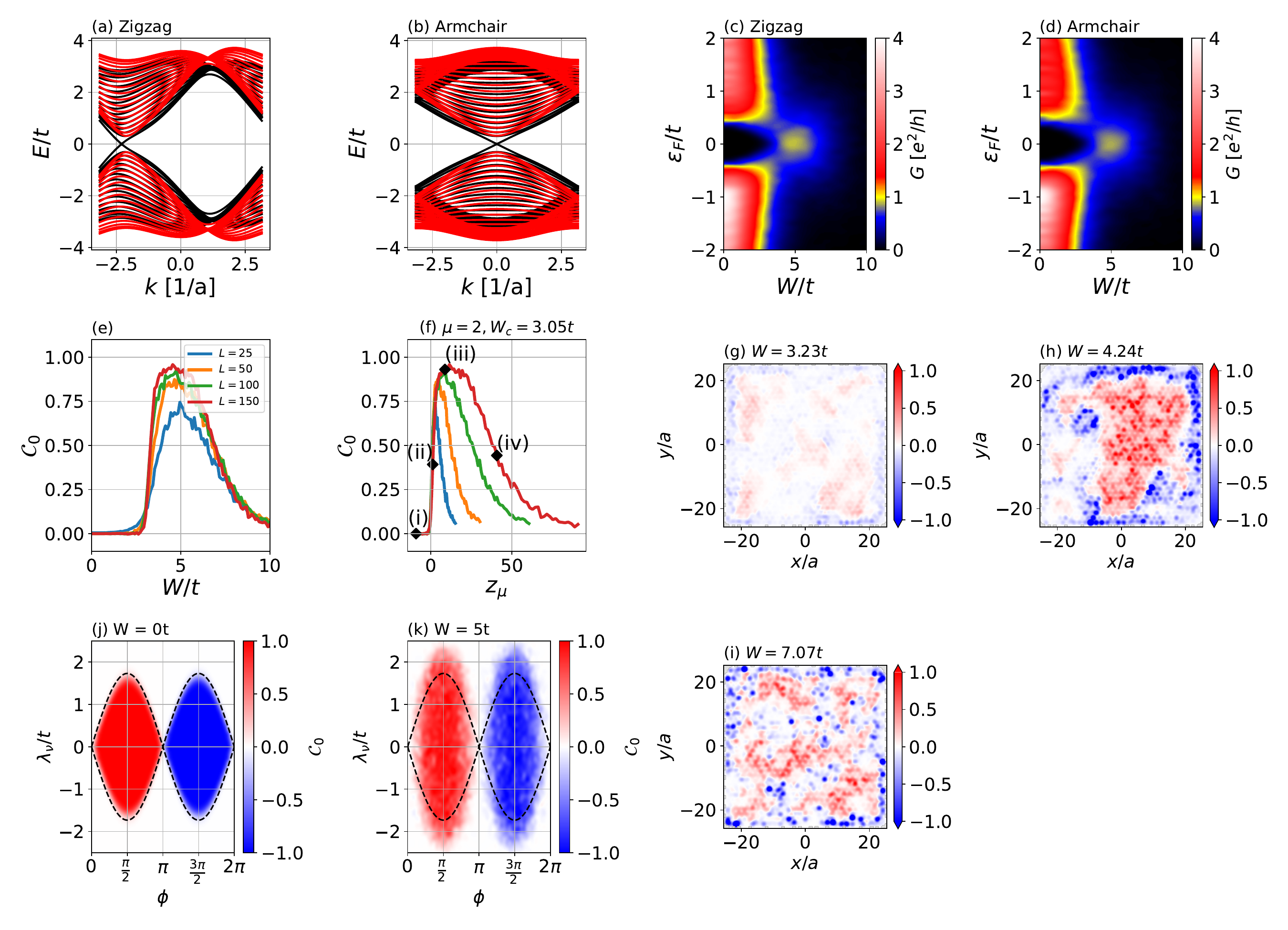}
    \caption{(a)-(b) Band structures for Haldane system in the trivial regime ($M = 2t$) in red, and non-trivial regime ($M = 1.5t$) in black, for (a) zigzag and (b) armchair ribbons. (c)-(d) Conductance $G$, averaged over 100 disorder configurations, as a function of the disorder intensity $W$ and chemical potential $\varepsilon_F$ for the Haldane (c) zigzag and (d) armchair ribbons of size $L_x = L_y = 25a$, with $\phi = \pi/2$ and $M = 2t$.
    (e) Local Chern marker, for $\phi = \pi/2$ and $M = 2t$, as a function of disorder intensity $W$, for $L_x = L_y = L$ averaged over 400 disorder realizations and 1000 Chebyshev moments in the KPM expansion. (f) Renormalization of the curves in (e) showing a universal superposition in the region of TAI after the scaling $z_\mu = (W^\mu - W_c^\mu) L/ (L_{\rm max} t^\mu)$, where the denominator $L_{\rm max} t^\mu$ is introduced to keep $z_\mu$ dimensionless.
    (g)-(i) Spatial distribution of the local Chern marker at all sites of the Haldane system with $L_x = L_y = 25a$, using $M = 2t$, $\phi = \pi/2$, and different values of $W= 3.23t$, $W= 4.24t$, and $W= 7.07t$ respectively. (j)-(k) Topological phase diagrams drawn from the local Chern marker ($\mathcal{C}_{0}$) as a function of the phase $\phi$ and mass $M/t$, without disorder (j) and with disorder $W= 5t$ (k).}
    \label{Haldane_results}
\end{figure*}

The band structure for the Haldane model is illustrated in Figs.~\ref{Haldane_results}(a) and \ref{Haldane_results}(b), which show the non-trivial (black lines) and trivial (red lines) energy bands for the zigzag and armchair ribbons, respectively. 
Figures~\ref{Haldane_results}(c) and \ref{Haldane_results}(d) show the conductance $G$ as a function of the chemical potential $\varepsilon_F$ and disorder intensity $W$ with $M = 2t$ and $\phi = \pi/2$ for zigzag and armchair ribbons, respectively. 
For $W = 0$ the system is in a trivial regime and the conductance vanishes in the gap around $\varepsilon_F = 0$. 
However, as the disorder intensity increases, a finite conductance $G \approx G_0$, where $G_0= e^2/h$, emerges within the gap region, as represented by the yellow region around $\varepsilon_F=0$, which is characteristic of topological Anderson insulators. 
Notably, the size of this region in parameter space is much smaller than the one we find for the BHZ model calculations in the main text, owing to the narrower bandwidth in the Haldane model.

The topological phase transition can be seen in Fig.~\ref{Haldane_results}(e), which shows the local Chern marker at the central cell of the system, averaged over 400 disorder configurations, as a function of $W$ for systems of various sizes $L$. 
These curves do not reach the full plateau $\mathcal{C}_0 = 1$ of the non-trivial regime, since the onset of Anderson localization occurs at small values of $W$ due to the narrow bandwidth of the Haldane model. 
Nevertheless, the scaling behavior remains equivalent to that described in the main text. 
Figure~\ref{Haldane_results}(f) shows the curves from Fig.~\ref{Haldane_results}(e) renormalized through the scaling $W \rightarrow (W^\mu-W_c^\mu)/W_{\rm max}^\mu \cdot (L/L_{\rm max})$, with a critical exponent of $\mu \approx 2$ and critical disorder of $W_c \approx 3.05t$. 

Figures~\ref{Haldane_results}(g), \ref{Haldane_results}(h), and \ref{Haldane_results}(k) show a color map of the local Chern marker over the system sites $C_0({\bm r})$ for disorder intensities of $W= 3.23t$, $W= 4.24t$, and $W= 7.07t$, respectively. 
These disorder strengths correspond to points (ii), (iii), and (iv) indicated in Fig.~\ref{Haldane_results}(f). 
For a small disorder intensity, Fig.~\ref{Haldane_results}(g), the local Chern marker at the center of the system is $\mathcal{C}({\bm R})\approx 0.5$ with small fluctuations, indicating a transition between the trivial and topological regimes. 
In contrast, Fig.~\ref{Haldane_results}(h) shows that the local Chern marker is $\mathcal{C} ({\bm R})\approx 1$ over most of the central region, resembling a bulk topological regime. 
Finally, Figure~\ref{Haldane_results}(i) shows that strong disorder drives the system back into the trivial regime, due to the emergence of Anderson localization.

In Fig.~\ref{Haldane_results}(j), we plot the phase diagram of the pristine system using the local Chern marker in the center of the system. The topological region depicted by the local Chern marker matches exactly the outlined region set by the dashed curve. In contrast, for a disordered system with disorder intensity $W=5t$, the phase diagram of Fig.~\ref{Haldane_results}(k) shows that the Anderson disorder extends the topological region beyond the dashed lines.

%%%%%%%%%%%%%%%%%%%%%%%%%%%%%%%%%%%%%%
\subsection{Kane-Mele model}
\label{sec:KaneMele}

\begin{figure*}[ht!]
    \includegraphics[width=\textwidth]{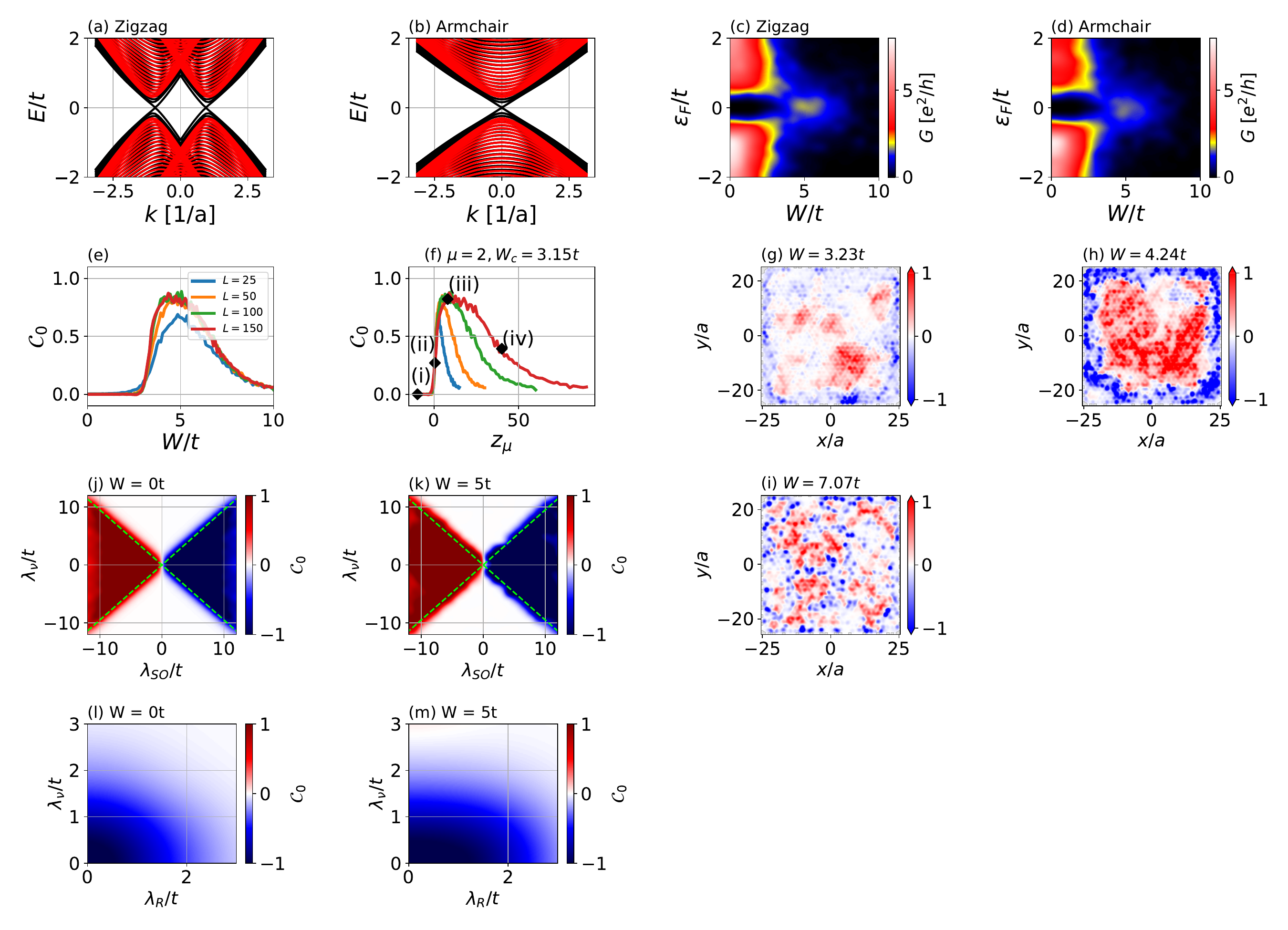}
    \caption{Band structures for Kane-Mele system in the trivial regime ($\lambda_{\nu} = 1.85t$) in red, and non-trivial regime ($\lambda_{\nu} = 1.45t$) in black, for (a) zigzag and (b) armchair ribbons, using $\lambda_{R} = 0$, $\lambda_{SO} = 0.3t$. 
    Conductance, averaged over 100 disorder configurations, as a function of the disorder intensity $W$ and chemical potential $\varepsilon_F$ for the Kane-Mele system of size $L_x = L_y = 25a$, $\lambda_{R} = 0$, $\lambda_{SO} = 0.3t$, and $\lambda_{\nu} = 1.85t$, for (c) zigzag and (d) armchair ribbons. 
    (e) Local Chern marker, for $\lambda_{R} = 0$ and $\lambda_{\nu} = 0.3t$, as a function of disorder intensity $W$ and various $L_x = L_y = L$, averaged over 400 disorder realizations and 1000 Chebyshev moments in the KPM expansion.
    (f) Same curves as in panel (e) scaled according to 
    $z_\mu = (W^\mu - W_c^\mu) L/ (L_{\rm max} t^\mu)$, where the denominator $L_{\rm max} t^\mu$ is introduced to keep $z_\mu$ dimensionless,
    showing a universal behavior in the TAI region. 
    (g)-(i) Local Chern markers ${\cal C}(\bm{R})$ as a function of the positions of the Kane-Mele system cells with $L_x = L_y = 50a$ for $\lambda_{\nu} = 1.85t, \lambda_{R} = 0,$ and $\lambda_{SO} = 0.3t$ and different values of $W= 3.23t$, $W= 4.24t$, and $W= 7.07t$ respectively. (j)-(k) Topological phase diagrams drawn from the local Chern marker as a function of spin-orbit coupling $\lambda_{\rm SO}/t$ and mass parameter $\lambda_{\nu}/t$, using $\lambda_{R} = 0.05t$, and $L_x = L_y = 15 \sqrt{3}a$, without disorder (j) and with disorder $W = 5t$ (k). (l)-(m) Topological phase diagrams drawn from the local Chern marker as a function of spin-orbit coupling $\lambda_{R}/t$, and mass parameter $\lambda_{\nu}/t$, using $\lambda_{SO} = 0.3t$ and $L_x = L_y = 15 \sqrt{3}a$, without disorder (l) and with disorder $W = 5t$ (m).}
    \label{KaneMele_results}
\end{figure*}

The Kane-Mele Hamiltonian \cite{KaneMele2005} reads
\begin{multline}
    H_{\rm KM} = t \sum_{\expval{ij}} c_i^\dagger c_j
    + \lambda_\nu \sum_i \xi_i c_i^\dagger c_i
    + i\lambda_{\rm SO} \sum_{\expval{\expval{ij}}} \nu_{ij} c_i^\dagger \sigma_z c_j 
    \\
    i\lambda_{\rm R} \sum_{\expval{ij}} c_i^\dagger (\bm{\sigma}\times \hat{\bm{d}}_{ij})_z c_j 
\end{multline}
where $i$ and $j$ label the sites of a honeycomb lattice and $\expval{ij}$ restrict the double sum to run over pairs of the nearest-neighbors sites, whereas $\expval{\expval{ij}}$ constraints the sum to next-nearest neighbors sites.
The operators $c_i^\dagger = (c_{i \uparrow}^\dagger, c_{i \downarrow}^\dagger)$ and $c_i = (c_{i \uparrow}, c_{i \downarrow})^{T}$ create and annihilate (spin-full) electrons at the site $i$ and
${\bm \sigma} = (\sigma_x, \sigma_y, \sigma_z)$ stands for the Pauli matrices acting on the spin subspace.
The term $\lambda_{\nu}$, corresponds to the sublattice potential, with $\xi_i = 1$ for the A sublattice and $\xi_i = -1$ for the B sublattice.
The spin-orbit interaction is described by the two-last terms of the Hamiltonian.
The coupling constants $\lambda_{SO}$ and $\lambda_{R}$ stand, respectively, for the intrinsic and the Rashba spin-orbit coupling (SOC). 
The intrinsic SOC has a phase determined by $\nu_{ij} = \frac{2}{\sqrt{3}}(\bm{d}_{i} \times \bm{d}_{j})_{z} = \pm 1$, where $\bm{d}_i$ and $\bm{d}_j$ are unit vectors along the two bonds the electron goes through when hopping from site $j$ to site $i$. For the Rashba SOC, $\hat{\bm{d}}_{ij} = \bm{d}_{ij}/d_{ij}$ is the unit vector pointing from site $j$ to $i$.

In the following we study $H_{\rm KM}$ in the presence of Anderson onsite disorder, namely,
\begin{equation}
    H = H_{\rm KM} + \sum_i \varepsilon_i c_i^\dagger c_i^{} \;,
\end{equation}
where $\varepsilon_i$ is randomly chosen from a uniform distribution within the interval $[-W/2, W/2]$. 
We express all energies in units of the hopping term $t$ and analyze the interplay between $\lambda_{\rm SO}, \lambda_\nu, \lambda_{\rm R}$ and $W$.

The band structure of the Kane-Mele model is depicted in Figs.~\ref{KaneMele_results}(a) and \ref{KaneMele_results}(b), which show the non-trivial (black lines, with $\lambda_{\nu}=1.45t$) and trivial (red lines, with $\lambda_{\nu}=1.85t$) bands for zigzag and armchair ribbons, respectively. 
Similar to the Haldane model, the Kane-Mele model features a narrow bandwidth compared to the BHZ model. Consequently, the onset of the standard Anderson localization phase occurs already at small values of $W/t$ and limits the range of parameters where the TAI phase can be seen.

Figures \ref{KaneMele_results}(c) and \ref{KaneMele_results}(d) show the average Landauer conductance $G$ for 100
realizations as a function of the chemical potential $\varepsilon_F$ and disorder intensity $W$ for the Kane-Mele model for zigzag and armchair ribbons. 
In these simulations, the mass value is fixed at $\lambda_{\nu} = 1.85t$, with $\lambda_{R} = 0$ and $\lambda_{SO} = 0.3t$, and the system size is $L_x = L_y =25a$. 
For weak disorder, $W/t \alt 1$, the conductance vanishes around $\varepsilon_F = 0$ due to the gap seen in the trivial regime. As disorder intensity increases, a domain of quantized conductance $G \approx G_0$ emerges within the gap region, corresponding to the yellow region around $\varepsilon_F=0$, characteristic of topological Anderson insulators. 
However, the size in the parameter space of this quantized conductance region is smaller than that of the BHZ model, owing to the narrower bandwidth. 

Figure \ref{KaneMele_results}(e) illustrates the topological phase transition in terms of the local spin Chern marker ${\cal C}_0$, averaged over 400 disorder configurations, at the center of the system as a function of disorder intensity $W$. 
Here, the Rashba spin-orbit coupling is absent, $\lambda_R = 0$, which allows the use of spin Chern number since $[H, \sigma_z] = 0$. In this case, the local marker is calculated using $\hat{U} = \sigma_z$ (see Eq.~3 in the main text).
While there is a transition to the non-trivial phase, it does not reach the fully non-trivial phase where ${\cal C}_0 = 1$, due to the small bandwidth of the model which leads to the emergence of the standard Anderson localization phase at small $W$.
The scaling analysis, see Fig.~\ref{KaneMele_results}(f), gives a critical exponent $\mu \approx 2$, which is consistent with the ones we obtain for the BHZ (main text) and Haldane models (SM above). 

Figures~\ref{KaneMele_results}(g), \ref{KaneMele_results}(h) and \ref{KaneMele_results}(i) show the full color map of the local Chern marker over the system sites 
${\cal C}({\bm R})$ for disorder intensities of $W= 3.23t$, $W= 4.24t$, and $W= 7.07t$, respectively. 
These intensities correspond to the points (ii), (iii), and (iv) in Fig.~\ref{KaneMele_results}(f). 
For $W= 3.23t$, Fig.~\ref{KaneMele_results}(g), the local Chern marker intensity is $\mathcal{C}({\bm R}) \approx
0.5$ with small fluctuations, that is a disorder strength domain corresponding to the transition between trivial and topological regimes. 
In contrast, in Fig.~\ref{KaneMele_results}(h) the local Chern marker is $\mathcal{C}({\bm R}) \approx 1$ over almost the entire central region, resembling a bulk topological regime. 
Figure~\ref{KaneMele_results}(i) shows that an increasing disorder strength drives the system back into the trivial regime with the emergence of Anderson localization.

Next, we use the local spin-Chern marker $\mathcal{C}_0$ to draw phase diagrams for the Kane-Mele model with $\lambda_R \neq 0$. In this case $[H, \sigma_z] \neq 0$ and the spin is not a good quantum number. Nevertheless, as shown by Refs.~\cite{Ezawa2013, Ezawa2014, Zhu2019}, the spin-Chern number for finite $\lambda_R$ is quasi-quantized and remains a reliable witness of the topological phase.

Figure~\ref{KaneMele_results}(j), shows the phase diagram of the pristine system using the local spin Chern marker as a function of the mass $\lambda_{\nu}/t$ and the spin-orbit coupling $\lambda_{\rm SO}$. 
The topological region is bounded by the dashed lines. 
In contrast, in the disordered case, illustrated by Fig.~\ref{KaneMele_results}(k), in which $W = 5t$, the Anderson disorder extends the topological region beyond the dashed line.

Similarly, we present a diagram of the local spin Chern marker as a function of mass $\lambda_{\nu}/t$ and $\lambda_{R}$ for both the pristine case, Fig.~\ref{KaneMele_results}(l), and the disordered one, Fig.~\ref{KaneMele_results}(m) with $W = 5t$.
The simulations indicate that in disordered systems the topological parameter space region is expanded with respect to the pristine limit, especially for larger values of $\lambda_{\rm R}$.
This conclusion is in line with the findings reported in Ref.~\cite{Orth2016}.

% to help adjust the figure placement
% \clearpage

\bibliography{main}